\documentclass[%
 reprint,
 amsmath,amssymb,
 prapplied,
]{revtex4-1}

\usepackage[utf8]{inputenc}
\usepackage{graphicx}% Include figure files
\usepackage{dcolumn}% Align table columns on decimal point
\usepackage{bm}% bold math
\usepackage{hyperref}% add hypertext capabilities
\usepackage{xspace} % for spaces in math expressions
\usepackage{mathrsfs} % for math fonts
\usepackage[inline]{enumitem} % for in line enumeration
\usepackage{color}
\usepackage{mathptmx}
\usepackage{amsmath} %allows to use align package
\usepackage{booktabs}

\usepackage[version-1-compatibility]{siunitx}
\DeclareMathAlphabet{\pazocal}{OMS}{zplm}{m}{n} %for fancy math symbols

\newcommand{\HH}{\ensuremath{\pazocal{H}}\xspace}
\newcommand{\smmu}{\ensuremath{\mu_{\mathrm{s}}}\xspace}
\newcommand{\smJij}{\ensuremath{J_{ij}}\xspace}

\newcommand{\smKu}{\ensuremath{k_{\mathrm{u}}}\xspace}

\newcommand{\Heffvec}{\ensuremath{\vec{ \mathrm{H} }_{\mathrm{eff}}}\xspace}

\newcommand{\Hthvec}{\ensuremath{\vec{ \mathrm{H} }_{\mathrm{th}}}\xspace}
\newcommand{\Happvec}{\ensuremath{\vec{ \mathrm{H} }_{\mathrm{app}}}\xspace}
\newcommand{\Hanivec}{\ensuremath{\vec{ \mathrm{H} }_{\mathrm{ani}}}\xspace}
\newcommand{\vampire}{\textsc{vampire}\xspace}
\newcommand{\Tc}{\ensuremath{T_{\mathrm{c}}}\xspace}
\newcommand{\kB}{\ensuremath{k_{\mathrm{B}}}\xspace}
\newcommand{\muB}{\ensuremath{\mu_{\mathrm{B}}}\xspace}
\newcommand{\Mags}{\ensuremath{M_{\mathrm{s}}}\xspace}

\newcommand{\Ku}{\ensuremath{K_{\mathrm{u}}}\xspace}

\newcommand{\muzero}{\ensuremath{\mu_0}\xspace}
\newcommand{\BcSW}{\ensuremath{B_{\mathrm{c}}^{\mathrm{SW}}}\xspace}

\newcommand{\BcLin}{\ensuremath{B_{\mathrm{c}}^{\mathrm{l}}}\xspace}
\newcommand{\zetaad}{\ensuremath{\vec{\zeta}_{\mathrm{ad}}}\xspace}
\newcommand{\zetaperp}{\ensuremath{\vec{\zeta}_{\bot}}\xspace}
\newcommand{\alphapara}{\ensuremath{\alpha_{\parallel}}\xspace}
\newcommand{\alphaperp}{\ensuremath{\alpha_{\bot}}\xspace}
\newcommand{\mvec}{\ensuremath{\vec{\mathrm{m}}}\xspace}
\newcommand{\Hintragrainvec}{\ensuremath{\vec{ \mathrm{H} }_{\mathrm{intragrain}}}\xspace}

\newcommand{\Chiperp}{\ensuremath{\Tilde{\chi}_{\perp}}\xspace}
\newcommand{\Chipara}{\ensuremath{\Tilde{\chi}_{\parallel}}\xspace}
\newcommand{\Chiparaperp}{\ensuremath{\Tilde{\chi}_{\parallel,\perp}}\xspace}
\newcommand{\Chilong}{\ensuremath{\Tilde{\chi}_{l}}\xspace}

\newcommand{\Tmin}{\ensuremath{T_{\mathrm{min}}}\xspace}

\newcommand{\Tpeak}{\ensuremath{T_{\mathrm{peak}}}\xspace}
\newcommand{\tpulse}{\ensuremath{t_{\mathrm{pulse}}}\xspace}

\newcommand{\Happ}{\ensuremath{H_{\mathrm{app}}}\xspace}
\newcommand{\Hmax}{\ensuremath{H_{\mathrm{max}}}\xspace}

\newcommand{\pmax}{\ensuremath{p_{\mathrm{max}}}\xspace}
\newcommand{\etal}{\textit{et al}\xspace}

\usepackage[normalem]{ulem}

\begin{document}

\preprint{APS/123-QED}

\title{Magnetisation dynamics of granular HAMR media by means of a multiscale model}

\author{A. Meo}
    \email{andrea.m@msu.ac.th}
    \affiliation{Department of Physics, Mahasarakham University, Mahasarakham, Thailand}
\author{W. Pantasri}
    \affiliation{Department of Physics, Mahasarakham University, Mahasarakham, Thailand}
\author{W. Daeng-am}
    \affiliation{Department of Physics, Mahasarakham University, Mahasarakham, Thailand}
\author{S. E. Rannala}
    \affiliation{Department of physics, University of York, York, UK}
\author{S. I. Ruta}
    \affiliation{Department of physics, University of York, York, UK}
\author{R. W. Chantrell}
    \affiliation{Department of physics, University of York, York, UK}
\author{P. Chureemart}
    \affiliation{Department of Physics, Mahasarakham University, Mahasarakham, Thailand}
\author{J. Chureemart}
    \email{jessada.c@msu.ac.th}
    \affiliation{Department of Physics, Mahasarakham University, Mahasarakham, Thailand}

% \date{\today}

\begin{abstract}
Heat assisted magnetic recording (HAMR) technology represents the most promising candidate to replace the current perpendicular recording paradigm to achieve higher storage densities. 
To better understand HAMR dynamics in granular media we need to describe accurately the magnetisation dynamics up to temperatures close to the Curie point.
To this end we propose a multiscale approach based on the micromagnetic Landau-Lifshitz-Bloch (LLB) equation of motion parametrised using atomistic calculations.
The LLB formalism describes the magnetisation dynamics at finite temperature and allows to efficiently simulate large system sizes and long time scales. 
Atomistic simulations provide the required temperature dependent input quantities for the LLB equation, such as the equilibrium magnetisation and the anisotropy, and can be used to capture the detailed magnetisation dynamics.
The multiscale approach makes possible to overcome the computational limitations of atomistic models in dealing with large systems, such as a recording track, while incorporating the basic physics of the HAMR process.
We investigate the magnetisation dynamics of a single FePt grain as function of the properties of the temperature profile and applied field and test the micromagnetic results against atomistic calculations. 
Our results prove the appropriateness and potential of the approach proposed here where the granular model is able to reproduce the atomistic simulations and capture the main properties of a HAMR medium.
\end{abstract}

%Use showkeys class option if keyword

\maketitle

\section{Introduction}
The continuous increase in the virtual data generated by computers and mobile devices is pushing the limit of the current storage technology and alternatives are required. 
Current hard disk drives are able to reach areal storage densities up to about \SI{1}{Tb in^{-2}} \cite{Rottmayer2006,Weller2016} with perpendicular magnetic recording (PMR) technology, but face limitations to increase it beyond this point due to the so called ``magnetic recording trilemma'' \cite{Evans2012a}: 
to further increase the areal storage density of recording media, smaller grains are needed;
these grains need to have a high magnetic anisotropy \cite{Chureemart2017} to be thermally stable;
to write these high anisotropy grains, large head fields are required and these cannot be provided by a conventional write head. 
Heat assisted magnetic recording (HAMR) \cite{Rottmayer2006} represents the most promising alternative to conventional magnetic recording. 
HAMR technology exploits the fact that the magnetic anisotropy of a ferromagnetic material decreases with temperature as this approaches the Curie point (\Tc). 
By heating the magnetic layer with a short and intense laser pulse to temperatures around \Tc, the data can be written using a weaker magnetic field without affecting the data stability. 
The temperature assist makes possible the use of grains with larger magnetic anisotropy, therefore allowing for smaller grain diameters. 
These improvements have made it possible to obtain a storage densities of \SI{1.4}{Tb in^{-2}}, as recently demonstrated by Seagate \cite{Ju2015}.

Despite HAMR being proposed and investigated for around 15 years, a complete understanding of the functioning of these devices necessary for the introduction into the market is lacking. 
Moreover, engineering the medium by combining ferromagnets with different properties and improving the head design can yield further increase in the storage density without compromising the reliability of the device. 
In order for HAMR to be reliable, it is necessary that grains adjacent to the targeted region are not affected during the writing process, as this could cause the undesired writing of such grains yielding errors in the reading of the signal. 
We aim to investigate on a theoretical and computational level the effects of temperature profile and thermal gradient on the magnetisation dynamics and writing process of a realistic HAMR medium to be able to suggest improved design of the magnetic stack and writing head. 
We utilise a multiscale model of a granular HAMR medium where an atomistic spin model is combined with a micromagnetic (granular) approach. 
The atomistic approach is primarily employed to parametrise the main magnetic properties of the magnetic materials, such as magnetisation, magnetic anisotropy, exchange coupling and damping constant. 
This information is then used as input into the macroscopic spin (granular) model to investigate the magnetisation dynamics in HAMR. 
The detailed mechanism of the magnetisation reversal is also simulated by means of atomistic simulations, although such a study is limited to relatively small regions due to the heavy computational requirements. 
The comparison between the results obtained with the atomistic approach and the granular model allows us to validate our multiscale approach and  provide extremely useful insights about the HAMR dynamics.

\section{Model}
\subsection{Atomistic model}
In the atomistic spin model one assumes that the magnetic moment can be localised on each atom, an approximation that works for the magnetic materials of interest in this work. 
Here the atomistic simulations are performed using the freely distributed software package \vampire \cite{vampire}, where the interactions contributing to the internal energy are given by the following extended Heisenberg Hamiltonian \cite{vampire-rev}: 
\begin{equation}
\HH = - \sum_{i < j} J_{ij} \vec{S}_i\cdot \vec{S}_j - \sum_{i} \smKu^i (\vec{S}_i\cdot \hat{e})^2 - \muzero\sum_i \smmu^i \vec{S}_i \cdot \Happvec \,.
  \label{eq:Hamiltonian}
\end{equation}
$J_{ij}$ is the exchange coupling constant for the interaction between the spins on site $i$ ($\vec{S}_i$) and $j$ ($\vec{S}_j$), $\smKu^i$ is the on-site uniaxial energy constant on site $i$ along the easy-axis $\hat{e}$, $\smmu^i$ is the atomic spin moment on the atomic site $i$ in units of \muB, \muzero is the permeability constant and \Happvec is the external applied field.
The first term on the right hand side (RHS) of Eqn.~\ref{eq:Hamiltonian} represents the exchange coupling, the second the magnetic anisotropy energy and the third the interaction with an external magnetic field.
The dynamics of each individual spin is obtained by integrating the Landau-Lifshitz-Gilbert (LLG) equation of motion  \cite{vampire-rev}: 
\begin{align}
\frac{\partial \vec{S}_i}{\partial t} =& - \frac{\gamma_e}{1+\lambda^2}[\vec{S}_i \times \Heffvec^i + \lambda\vec{S}_i \times (\vec{S}_i \times \Heffvec^i)] \, .
\label{eq:LLG}
\end{align}
$\gamma_e = \SI{1.761e11}{T^{-1} s^{-1}}$ is the electron gyromagnetic ratio, $\lambda$ controls the damping and represents the coupling of spins to a heat bath through which energy can be transferred into and out of the spin system.
$\Heffvec^i$ is the effective field acting on each spin obtained by differentiating the Hamiltonian (Eqn.~\ref{eq:Hamiltonian}) with respect to the atomic spin moment and accounts for the interactions within the system. 
Finite temperature effects are included under the assumption that the thermal fluctuations are non-correlated and hence can be described by a white noise term.
This is expressed as a Gaussian distribution in 3 dimensions whose first and second statistical moments of the distribution are:
\begin{align}
\label{eq:thermal-field-first-moment}
\left\langle \xi_{i\alpha} (t) \right\rangle &= 0 , \\
\label{eq:thermal-field-second-moment}
\left\langle \xi_{i a} (t) \xi_{j b} (t\prime) \right\rangle &= \frac{2 \lambda \kB T}{\smmu \gamma} \delta_{ij} \delta_{ab} \delta(t-t^\prime) ,
\end{align}
where $i$, $j$ label spins on the respective sites, $a,b = x,y,z$ are the vector component of $\vec{\xi}$ in Cartesian coordinates, $t, t^\prime$  are the time at which the Gaussian fluctuations are evaluated, $T$ is the temperature, $\delta_{ij}$ and $\delta_{ab}$ are Kronecker delta and $\delta(t-t^\prime)$ is the delta function.
Eqn. \ref{eq:thermal-field-first-moment} represents the average of the random field, whilst Eqn. \ref{eq:thermal-field-second-moment} gives the variance of the field, which is a measure of the strength of its fluctuations. 
The thermal contribution can be added to $\Heffvec^i$:
\begin{align}
\Heffvec^i =& -\frac{1}{\smmu^i} \frac{\partial \HH}{\partial \vec{S}_i} +  \Hthvec^i \, .
\label{eq:LLG-Hthermal}
\end{align}

\subsection{Granular model}
In our granular micromagnetic approach the magnetic medium is comprised of grains to each of which a microspin \mvec is associated.
Since HAMR devices involve the heating via a laser pulse of the magnetic medium close or up to \Tc, a micromagnetic model based on the LLG dynamics is not the most appropriate choice, as this considers the length of the magnetisation constant. 
Garanin \cite{Garanin1997a} derived a macrospin equation of motion, the Landau-Lifshitz-Bloch (LLB equation which accounts for the longitudinal relaxation of the magnetisation. This is clearly important at elevated temperatures  and therefore this is the formalism used in this work.
Our micromagnetic simulations are based on the stochastic form of the LLB equation
implemented following the work of Evans \etal \cite{LLB-II}.
The LLB equation of motion for each grain $i$ reads:
\begin{widetext}
    \begin{equation}
        \label{eq:sLLB-II}
        \frac{\partial{\mvec^i}}{\partial{t}} = 
        \gamma_e \left( {\mvec^i} \times \Heffvec^i \right) - \frac{\gamma_e\alphapara}{{m^i}^2} \left(\mvec^i \cdot \Heffvec^i \right) \mvec^i + \frac{\gamma_e\alphaperp}{{m^i}^2} \left[ \mvec^i \times \left( \mvec^i \times \left(\Heffvec^i + \zetaperp\right) \right) \right] + \zetaad \,.
    \end{equation}
\end{widetext}
$\gamma_e$ is the electron gyromagnetic ratio, $\mvec^i$ is the reduced magnetisation of grain $i$ which represents the vector magnetisation $\vec{M}^i$ normalised by its equilibrium magnetisation \Mags and $m^i$ is the length of $\mvec^i$. 
The first and third terms on the RHS of Eqn. \ref{eq:sLLB-II} are the precessional and damping terms respectively for the transverse component of the magnetisation, as in Eqn. \ref{eq:LLG}, while the second and fourth terms are introduced to account for the reduction of the longitudinal component of the magnetisation with temperature.
\alphapara and \alphaperp are the longitudinal and transverse damping parameters given by:
\begin{equation}
    \label{eq:damping}
    \alpha_{\parallel} = \frac{2}{3}\frac{T}{T_c}\lambda\quad\text{ and }\quad \left \{
    \begin{aligned}
        \alpha_{\bot} = \lambda(1-\frac{T}{3T_c}), && \text{if } T \leq T_c \\
        \alpha_{\bot} = \alpha_{\parallel} = \frac{2}{3}\frac{T}{T_c}\lambda, && \text{otherwise.}
    \end{aligned} \right.
\end{equation}
In Eqn.\ref{eq:damping} $\lambda$ is the atomistic damping parameter that couples the spin system with the thermal bath, the same entering the LLG equation for the atomistic approach (Eqn. \ref{eq:LLG}).
$\zeta_{\bot}$ and $\zeta_{\mathrm{ad}}$ are the terms that account for the thermal fluctuations in the limit that these can be treated as white noise. 
The thermal fields are described by Gaussian functions with zero average and variance (proportional to the strength of the fluctuations), analogously to the atomistic approach:
\begin{equation}
    \label{eq:diff_coef}
    \begin{aligned}
        <\zeta_{ad}^{i}(t)\zeta_{ad}^{j}(t-t^\prime)> = \frac{2k_{B}T\alpha_{\parallel}}{\gamma MsV}\delta_{ij}\delta_{ab}\delta(t) \\
        <\zeta_{\bot}^{i}(t)\zeta_{\bot}^{j}(t-t^\prime)> = \frac{2k_{B}T(\alpha_{\bot}-\alpha_{\parallel})}{\gamma MsV\alpha_{\bot}^2}\delta_{ij}\delta_{ab}\delta(t) \,.
    \end{aligned}
\end{equation}
$\Heffvec^i$ is the effective field that acts on each grain $i$:
\begin{equation}
    \Heffvec= \Hanivec + \Hintragrainvec + \Happvec \, 
\end{equation}
The anisotropy field \Hanivec is described following Garanin's approach \cite{Garanin1997a}:
\begin{equation}
    \Hanivec = \left(m_x\hat{e}_x+m_y\hat{e}_y\right)/\Chiperp \,,
\end{equation} 
where $\hat{e}_{x,y}$ is the unit vector aligned along the $x,y$--directions, $m_{x,y}$ the reduced magnetisation components along $x,y$--axis and \Chiperp is the reduced perpendicular susceptibility, which gives the strength of the fluctuations of the components of the magnetisation transverse to the easy-axis and introduces the temperature dependence in \Hanivec. 
This expression for the anisotropy field reduces to $2K/\Mags$ at $T=\SI{0}{\kelvin}$.

The intragrain exchange field $\Hintragrainvec$ accounts for the exchange between the atoms within the grain $i$ controlling the length of the magnetisation and has the form:
\begin{equation}
    \label{eq:Intragrain}
    \Hintragrainvec = \left \{
    \begin{aligned}
            \frac{1}{2\Chipara}\bigg(1-\frac{m^2}{m^2_e}\bigg)\mvec, && \text{if } T \leq T_c \\
           -\frac{1}{\Chipara}\bigg(1+\frac{3}{5}\frac{\Tc}{T-T_c}m^2\bigg)\mvec, && \text{otherwise}
    \end{aligned} \right. 
\end{equation}
where $m$ is length of the grain reduced magnetisation \mvec, $m_e(T)$ is the equilibrium magnetisation and $\Chipara$ is the reduced parallel component of the susceptibility.
\Chipara represents the magnetisation fluctuations along the easy-axis and depends on temperature, as \Chiperp.
\Happvec represents the externally applied magnetic field used to reverse the direction of the magnetisation. 

\subsection{HAMR dynamics}
In this study we concentrate on individual grain dynamics and the atomistic parameterisation of the macrospin LLB equation. We consider a simple analogue of the HAMR process in which an external field \Hmax is applied to the region under the writing head. 
The laser pulse $T(t)$ is modelled as a temperature pulse with Gaussian profile in time $T(t)$ while it is uniform in space:
\begin{equation}
    \label{eq:total_temperature_profile}
    T(t) = \Tmin + \Big[\Tpeak-\Tmin\Big]F(t).
\end{equation}
where 
\begin{equation}
    \label{eq:temporal_temperature_profile}
    F(t) = \exp{\Bigg[-\Bigg(\frac{t-3\tpulse}{\tpulse}\Bigg)^2\Bigg]} 
\end{equation}
is a Gaussian in one dimension with standard deviation $\sqrt{2}\tpulse$ and hence the maximum temperature \Tpeak of the pulse is reached at 3\tpulse.
\Tmin is the temperature at which the system is left when no pulse is applied, usually room temperature.
We remark that the results presented in this work are aimed to prove the goodness of the proposed approach and that these represent initial findings on a simple system composed of single grain of the granular layer. 
As such, here we neglect the spatial dependence of the heat pulse and include only its time dependence.
More complex systems and dynamics are object of further studies and are not discussed in this work.

\section{Results}

\subsection{Multiscale parameterisation of granular medium model}
We consider a HAMR medium whose magnetic layer is composed of a single layer of identical, non-interacting grains comprised of fully chemically ordered tetragonally distorted fcc (fct) L1-0 FePt, where Fe and Pt occupy distinct planes.
In this phase FePt is characterised by a large magnetocrystalline anisotropy \cite{Weller2016} directed along the long axis of the grain ($z$-axis) that provides the required thermal stability to retain the data over 10 years, and relatively low \Tc around \SI{700}{K}.
Moreover, ordered L1-0 FePt exhibits long range exchange coupling and two-ion anisotropy energy, which represents an anisotropic exchange interaction \cite{Mryasov2005,Natalia_th,Ellis2015b}. 
We model FePt mapping the fct structure to a distorted sc crystal structure with lattice vectors in $x$, $y$ and $z$ directions $a_{0,x}=a_{0,y}=\SI{0.272}{nm}$, $a_{0,z}=\SI{0.385}{nm}$.
Mryasov \textit{et al} \cite{Mryasov2005} showed that the Pt moments are entirely induced by the Fe and can be replaced by substitution and an enhanced Fe moment of  \SI{3.23}{\muB}. This yields a saturation magnetisation \Mags of \SI{1.1e6}{J T^{-1} m^{-3}} as in bulk FePt \cite{Ellis2015b}. 
Here we use a simplified version of the Hamiltonian of ref \cite{Mryasov2005} in which atoms are assigned a uniaxial anisotropy energy with $\smKu=\SI{2.63e-22}{J atom^{-1}}$ and isotropic nearest-neighbours exchange coupling $\smJij=\SI{6.81e-21}{J link^{-1}}$.
We assume $\lambda=0.1$ for FePt, a value accepted normally for compounds including heavy elements such as Pt and in agreement with the value reported in \cite{Becker2014}.
Table \ref{table:HAMR_equilibrium_parameters_FePt}  summarises the material parameters used in our work.
\begin{table}[!h]
    \caption{Simulation parameters for the investigated systems.}
    \centering
    \begin{tabular}{ccc}
        \toprule
        \toprule
  			       &   FePt     		&   Unit            \\
        \midrule
        \midrule
        \smJij      & \SI{6.81e-21}{}   &  J link$^{-1}$ \\ 
        \smKu       & \SI{2.63e-22}{}   &  J atom$^{-1}$ \\
        \smmu       & 3.63	            &  \muB          \\
        \Tc         & \SI{690}{}      &  \SI{}{K}         \\ 
        \Ku         & \SI{9.23}{}       &  \SI{}{\joule\per\cubic\metre}         \\
        \muzero\Mags& \SI{1.32}{}       &  T                \\
        $\lambda$    & 0.10              &                \\
        \bottomrule
        \bottomrule
    \end{tabular}
    \label{table:HAMR_equilibrium_parameters_FePt}
\end{table}

\begin{figure}[!tb]
    \centering
    \includegraphics[width=1.0\columnwidth]{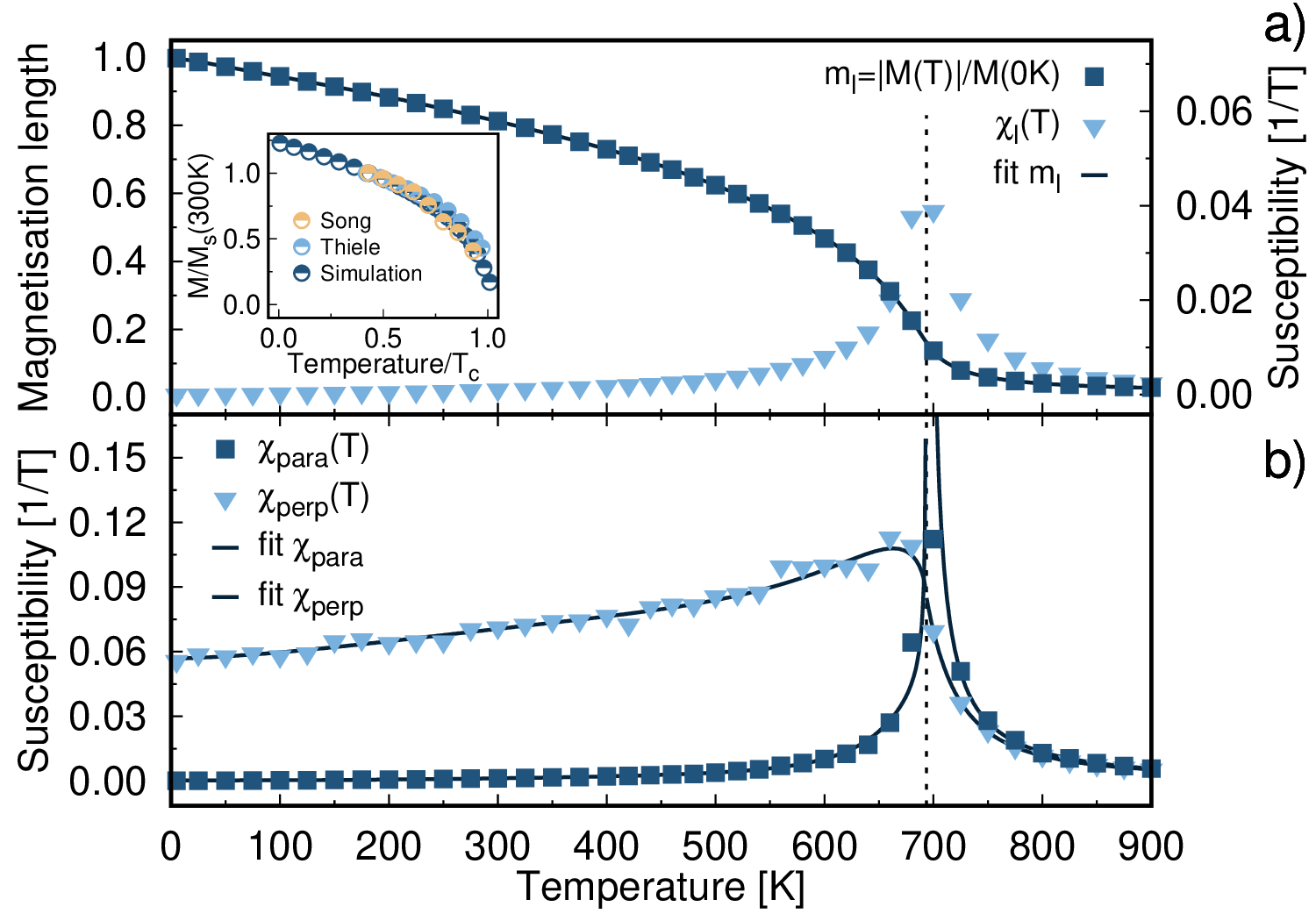}
    \caption{Temperature dependence of (a) reduced magnetisation length $m=|\vec{M}|/M(\SI{0}{\kelvin})$ and reduced longitudinal susceptibility \Chilong, (b) reduced parallel and perpendicular susceptibility \Chipara and \Chiperp, respectively, for a \SI{5 x 5 x 10}{nm} FePt grain. Black continuous lines are fit to the data according to Eqs.\ref{eq:m_eq_fit} and \ref{eq:susceptibility_fit}.
    The comparison between calculated (dark blue) and measured (light blue from Thiele\cite{thiele2002}, yellow from Song\cite{Song2017a}) magnetisation temperature dependence normalised by the saturation value at \SI{300}{\kelvin} is presented in the inset. Temperatures are normalised by the respective system Curie temperature \Tc.
    }
    \label{fig:M_K_X_vsTemp_FePt_5x5x10}
\end{figure}

We determine the temperature dependent equilibrium magnetisation and susceptibility components by performing time evolution of the magnetisation and by averaging over 100 repetitions. We integrate the spin system for \SI{50000}{} steps, after an equilibration of \SI{50000}{} to ensure good convergence of the results, with an integration step $dt=\SI{1}{\femto\second}$.
Classical spin dynamics yields a critical exponent of the magnetisation as function of \Tc around \SI{0.3}, a value close to what we obtain by fitting our simulations results assuming a bulk  behaviour.
Experimentally, the magnetisation shows a flatter trend at low temperature and a more critical behaviour close to \Tc.
We compare our calculated magnetisation temperature dependence by means of atomistic simulations with the experimental results obtained by Thiele \etal \cite{thiele2002} and Song and collaborators \cite{Song2017a}, presented as inset in Fig.~\ref{fig:M_K_X_vsTemp_FePt_5x5x10}. 
Despite different values of \Mags and \Tc for each system, we do not observe significant differences between simulations and experiments when we normalise the data with respect to $M(\SI{300}{\kelvin})$ and \Tc in the temperature range of interest.
We note that this agreement is obtained without  applying a temperature rescaling \cite{Evans2015} which maps classical spin simulations onto the experimental behaviour in the cases where quantum statistics dominate at low temperatures.  
This differs from assumptions in other works \cite{Ababei2019}.

The granular LLB model requires the temperature dependence of the magnetisation and that of the perpendicular and parallel susceptibilities as input parameters.
We obtain these quantities by performing atomistic simulations and fitting the data, as shown in Fig.~\ref{fig:M_K_X_vsTemp_FePt_5x5x10} for a \SI{5 x 5 x 10}{nm} FePt grain.
The temperature dependent magnetisation length $m(T)=M(T)/M(\SI{0}{\kelvin})$ is fitted using a polynomial expression in $(T-\Tc)/\Tc$, as discussed by \citet{Kazantseva2009}:
\begin{equation}
    \label{eq:m_eq_fit}
    m(T) =
    \left \{
    \begin{aligned}
        \sum_{i=0}^{9} A_i \left( \frac{\Tc-T}{\Tc} \right)^i   &+ A_{1/2} \left( \frac{\Tc-T}{\Tc} \right)^{\frac{1}{2}} &&, \text{if } T < T_c \\
        \bigg[\sum_{i=1}^{2} B_i\left( \frac{T-\Tc}{\Tc} \right)^i &+ A^{-1}_0 \bigg]^{-1}  &&, \text{otherwise.}
    \end{aligned} \right.
\end{equation}
This formulation of $m(T)$ allows to reproduce the finite-size effects captured by the atomistic spin dynamics simulations, see Fig.~\ref{fig:M_K_X_vsTemp_FePt_5x5x10}.
The susceptibility expresses the strength of the fluctuations of the magnetisation and, according to the spin fluctuation model, the components of the susceptibility can be obtained by the fluctuations of the same magnetisation components as follows \cite{Ellis2015a}:
\begin{equation}
    \Tilde{\chi}_{\mathrm{\alpha}} = 
    \frac{\smmu N}{\kB T}
    \left(
    \left\langle m_{\alpha}^2 \right\rangle - \left\langle m_{\alpha} \right\rangle^2
    \right) \, .
    \label{eq:reduced_susceptibility}
\end{equation}
Here $\left\langle m_{\alpha}\right\rangle$ is the ensemble average of the reduced magnetisation component $\alpha=x,y,z,l$, $N$ is the number of spins in the system with magnetic moment \smmu, $\kB=\SI{1.381e-23}{J K^{-1}}$ is the Boltzmann constant and $T$ the temperature.
$l$ is the length of the magnetisation, whereas $x,y,z$ are the spacial components  of the magnetisation.
\Chipara refers to the magnetisation component along the easy-axis direction, which is $z$ for our system, whereas \Chiperp describes the fluctuations of the magnetisation in the plane perpendicular to the easy-axis.
For \Chipara and \Chiperp fitting functions we use a similar approach to Ellis \cite{Ellis2015a}: 
\begin{equation}
    \label{eq:susceptibility_fit}
    \frac{1}{\Chiparaperp} = 
    \left \{
    \begin{aligned}
        \sum_{i=0}^{9} C_i \left( \frac{\Tc-T}{\Tc} \right)^i &+ C_{1/2} \left( \frac{\Tc-T}{\Tc} \right)^{\frac{1}{2}} &&, \text{if } T < T_c \\
        \sum_{i=0}^{4} D_i \left( \frac{T-\Tc}{\Tc} \right)^i &  &&, \text{otherwise.}
    \end{aligned} \right.
\end{equation}
where $C_i$ and $D_i$ and \Tc are fitting parameters.
Once all these parameters are determined, the granular model is fully parametrised regarding the material properties.

\subsection{Simulations of HAMR dynamics}
We simulate the magnetisation dynamics of a single \SI{5 x 5 x 10}{nm} FePt grain varying the peak temperature \Tpeak, length of the temperature pulse \tpulse and values of the magnitude of the applied field \Hmax =0.5 and \SI{1}{T}, repeating each simulation 100 times to ensure a large enough statistical ensemble. For these simulations we use a smaller integration step of \SI{0.1}{\femto\second} to ensure the convergence of the results.
\begin{figure}[!h]
    \centering
    \includegraphics[width=0.8\columnwidth]{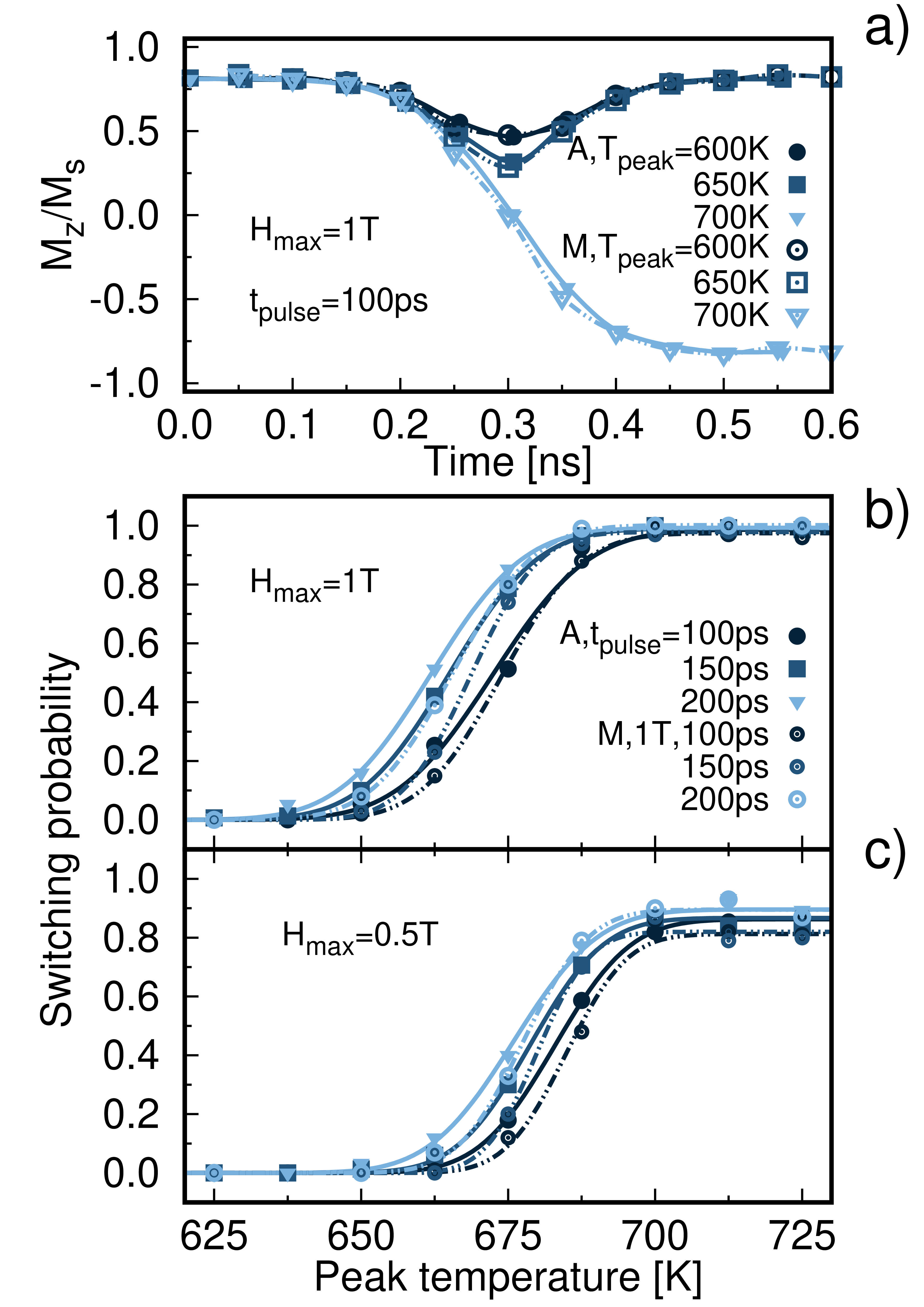}
    \caption{(a) Time evolution of the $z$-component of the magnetisation ($M_z/M_s$) of a \SI{5 x 5 x 10}{\nano\metre} hexagonal grain of FePt subjected to an external field of \SI{1}{T} and a temperature pulse \tpulse=\SI{100}{ps} as a function of temperature. 
    Comparison of atomistic (solid lines and filled symbols) and micromagnetic (dotted lines and empty symbols) switching probabilities for $\Hmax=\SI{1}{\tesla}$ (b) and \SI{0.5}{T} (c) as function of peak temperature for the same system.}
    \label{fig:M_vs_time_and_Pswitch_FePt}
\end{figure}
Fig.~\ref{fig:M_vs_time_and_Pswitch_FePt}(a) shows the time evolution of the $z$-component and length of the magnetisation subjected to a temperature pulse with $\tpulse=\SI{100}{ps}$ and different \Tpeak in an external field \Hmax of \SI{1.0}{T}.
For comparison, both atomistic and granular model calculations are shown.
After the temperature pulse reaches the maximum, where the magnetisation of the grain shrinks as $\Tpeak \sim \Tc$, and the temperature decreases, the external field can reverse the magnetisation.
The effect of peak temperature is further studied by computing the switching probability as function of peak temperature for different pulse duration. Results for applied field of \SIlist{1;0.5}{T} are shown in figure \ref{fig:M_vs_time_and_Pswitch_FePt}(b) and (c).
The application of a weaker \Hmax cannot succeed in switching the magnetisation due to the large thermal gradient of the pulse, which does not allow the individual spins within the grain to follow the field throughout the cooling process. 
The good agreement between the magnetisation dynamics obtained by performing atomistic simulations and by using the granular model is a proof that the latter incorporates the underlying thermal physics of the HAMR mechanism.

Further, we investigate the mechanism via which the magnetisation of a grain reverses during the writing process in HAMR systems. 
\begin{figure}[!h]
    \centering
    \includegraphics[width=1.025\columnwidth]{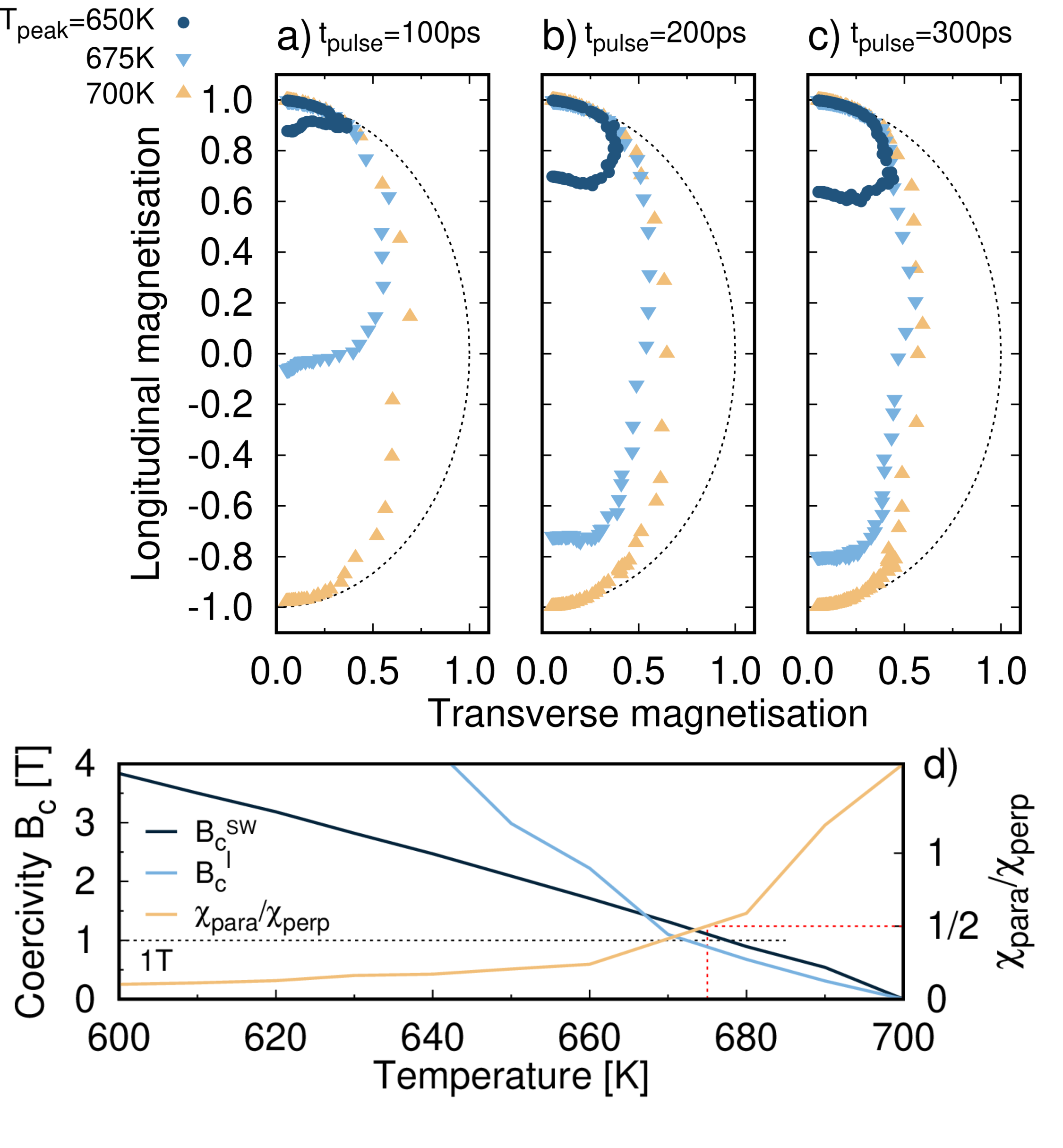}
    \caption{Plot of the average longitudinal magnetisation versus the transverse magnetisation of a \SI{5 x 5 x 10}{\nano\metre} hexagonal grain of FePt for 100 (a), 200 (b) and \SI{300}{\pico\second} (c) pulse times and for heat pulses reaching peak temperatures of \SIlist{650;675;700}{\kelvin} under the application of an external field $\Happ=\SI{1}{\tesla}$ calculated over 100 individual simulations. Dashed black lines show the circular reversal trajectory characteristic of a precessional dynamics.
    (d) Temperature dependence of coercivity for Stoner-Wohlfarth \BcSW and linear \BcLin reversal mechanism (left axis) and susceptibility ratio $\Chipara/\Chiperp$ (right axis) for the same FePt grains. Red dashed line marks the transition temperature for linear reversal following the work in Ref.~\cite{Kazantseva2009}, black dashed line marks the coercive field of \SI{1}{T}.
    }
    \label{fig:Mlong_vs_Mtran_atomvsmicro}
\end{figure}
We extract the average magnetisation along the easy-axis and perpendicular to it normalised by the average magnetisation length to account for the change in the length as the temperature changes. The former, longitudinal magnetisation, is $(M_z/\Mags)/(\sum_i M_l/N)$ and the latter, transverse magnetisation, is defined as $\sqrt{(M_x/\Mags)^2 + (M_x/\Mags)^2}/(\sum_i M_l/N)$.
The average runs over the 100 independent switching events mentioned above.
This is shown in Fig.~\ref{fig:Mlong_vs_Mtran_atomvsmicro} for different pulse lengths and peak temperatures with an applied field of \SI{1.0}{\tesla} for atomistic simulations. Results for micromagnetic simulations are not presented for the sake of clarity as they would overlap.
The dashed black line in panels a), b), c) depicts the circular reversal path characteristic of coherent precessional Stoner-Wohlfarth dynamics for a single domain particle, where to a reduction in the longitudinal magnetisation corresponds an increase of the transverse component. We can see that all the results start off following the circular trajectory until the transverse component reaches $\sim$0.3. As the magnetisation dynamics evolves,
the magnitude of the magnetisation clearly decreases on approaching the hard direction: the main characteristic of a transition to elliptical and linear reversal. The transition between these different regimes corresponds to a temperature of \SI{665}{K}. 
To understand the sudden change in the magnetisation behaviour at \SI{665}{K}, we look at the ratio of  susceptibilities \Chipara and \Chiperp for our system as a function of temperature, plotted as the yellow line in Fig.~\ref{fig:Mlong_vs_Mtran_atomvsmicro}(d). Since $1/\Chipara$ is proportional to the macroscopic longitudinal field of Eq.~\ref{eq:Intragrain} and $1/\Chipara$ represents the anisotropy field, the ratio $\Chipara/\Chiperp$ defines the transitions between reversal mechanisms, as discussed by Kazantseva \etal~\cite{Kazantseva2009}. Specifically, at low temperatures (for $\Chipara/\Chiperp<1/3$) the circular (coherent) mechanism is dominant. At this point elliptical reversal, involving a shrinking of the magnetisation along the hard direction, begins until $\Chipara/\Chiperp<1/2$ at which point the transverse magnetisation vanishes and linear reversal dominates.
Here we make a comparison of the characteristic switching fields to indicate the likely reversal mechanism for a given temperature range.
We extract the coercive field in case of Stoner-Wohlfarth dynamics ($\BcSW = 2K(T)/\Mags(T)$), presented in panel Fig.~\ref{fig:Mlong_vs_Mtran_atomvsmicro} panel d). We also show the switching field for linear reversal following Ref.~\cite{Kazantseva2009}.
The expectation, according to a simple transition from circular to elliptical and linear reversal suggests that at the highest temperature of \SI{700}{K} reversal should be completely linear. However, as shown in Fig.~\ref{fig:Mlong_vs_Mtran_atomvsmicro} panels a) to c) the reversal path we observe differs from the linear dynamics since the transverse component remains finite. We suggest that this is due to the timescale of the processes. According to Kazantseva \etal~\cite{Kazantseva2009} the characteristic timescale of reversal is strongly field dependent and in the temperature range of interest at \SI{1}{T} can be as much as many tens of picoseconds. As a result it is possible that at the rates of increase of temperature studied here the linear reversal mechanism is inaccessible. This further suggests a strong dependence of the reversal mechanism on the properties of the temperature pulse, such as duration and rate of increase. Therefore deeper analysis and investigations are required and will be object of further study. 

To better characterise the HAMR dynamics of our FePt grains, we register whether the grain magnetisation reverses and count one if the grain switches, zero otherwise.  By doing this, we build the switching probability of our system.  
We present in Fig.~\ref{fig:M_vs_time_and_Pswitch_FePt}(b) and (c) the switching probability as function of peak temperature \Tpeak comparing atomistic and micromagnetic simulations for $\Hmax=\SI{1.0}{\tesla}$ and $\Hmax=\SI{0.5}{\tesla}$, respectively.
Small differences can be observed when comparing the switching probabilities calculated using the two approaches. Consider first the case of an applied field of 1T shown in Fig ~\ref{fig:M_vs_time_and_Pswitch_FePt}. It can be seen that there is a small but systematic difference between the atomistic and macrospin model predictions with a shift of a few degrees between the respective probability curves.
We first observe that the range of temperatures we are considering is within \SI{80}{\kelvin} of \Tc, a critical regime for analytic approaches describing temperature dependent quantities. The LLB formalism was developed for bulk systems, and does not exhibit the reduced criticality of the finite size atomistic model simulations. Empirically, numerical parametrisation via Eqn.~\ref{eq:m_eq_fit} and Eqn.~\ref{eq:susceptibility_fit} is the simplest phenomenological approach to introduce finite size effects into the LLB formalism.
The small differences between the atomistic and macrospin model predictions suggest that the numerical parametrisation is a reasonable approach.

By exploiting the fact that each switching simulation is an independent event, we can treat it as a random variable and as such it is described by a normal distribution. 
The probability that the switching occurs is given by the cumulative distribution function.
By fitting the switching probability as function of peak temperature with the cumulative distribution function of a random variable we can extract the relevant parameters, such as the mean value $\mu$ and the width of the distribution $\sigma$. 
We express the cumulative distribution function following the discussion presented in \cite{Muthsam2019}:
\begin{equation}
    \label{eq:cumulative_func}
    \Phi(\mu,\sigma,\pmax) = 
    \frac{1}{2} \left[ 
    1 + \text{erf} \left( \frac{x - \mu}{\sqrt{2 \sigma^2}} \right)
    \right ] \pmax \, ,
\end{equation}
where $\text{erf} \left( x \right)$ is the error function and \pmax is the average maximum achievable switching probability.
$\sigma$ gives the steepness of the cumulative function and is a measure of the jitter noise, a parameter indicative of the maximum areal density achievable by the medium as it relates to the bit transitions. 
$\sigma \sim 1/(dP(T)/dT)$ and therefore steeper switching probability as function of temperature produce smaller jitter noise and are desirable. 
In addition, we can see from our results that $dP(T)/dT$ decreases with the magnitude of the applied field, in agreement with results presented in ref.\cite{Suess2015}, because the temperature window available to reverse the grain magnetisation reduces for a smaller applied field \Hmax. 
The maximum probability \pmax depends on the applied field via the temperature gradient of the switching field, hence higher \Hmax yields larger \pmax.
Because the total noise depends on both the field gradient and switching probability gradient with respect to temperature and these behave in opposite ways, a trade off is necessary to optimise HAMR media.
From the switching probability one can access the bit error rate (BER), as discussed by Vogler \etal \cite{Vogler2016a}. 
However, because of the low \pmax reached by the FePt system and keeping into consideration that the results shown here are for a system composed of uncoupled grains and a simple writing process where heat and field are applied uniformly to each grain is used, we do not compute the BER.

\begin{figure}[!h]
    \centering
    \includegraphics[width=1.0\columnwidth]{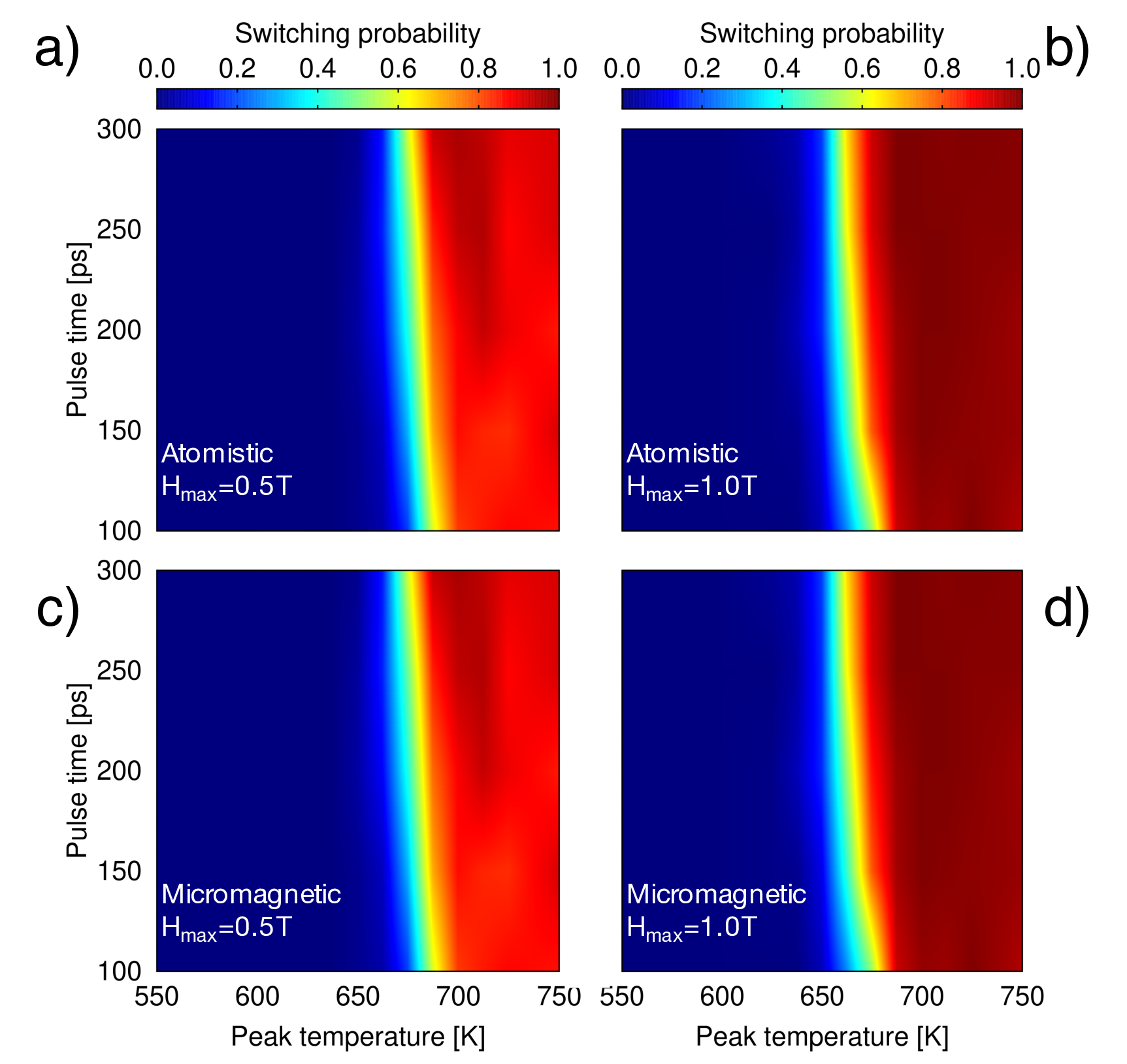}
    \caption{Plot of the switching probability (colour) for a single \SI{5 x 5 x 10}{\nano\metre} hexagonal grain of FePt as function of pulse length \tpulse and peak temperature \Tpeak for $\Happ=\SI{0.5}{\tesla}$ (left) and $\Happ=\SI{1.0}{\tesla}$ (right) comparing atomistic (top) and micromagnetic (bottom) simulations. 
    \Tpeak and \tpulse are varied with steps of \SI{12.5}{\kelvin} and \SI{50}{\pico\second}, respectively.
    }
    \label{fig:phase_plot_FePt_atomistic_and_granular}
\end{figure}
We combine the temperature and time dependence of the switching probability in phase plots showing the switching probability (colour) as function of the peak temperature \Tpeak and pulse time \tpulse for $\Hmax=\SIlist{0.5;1.0}{T}$, with steps of \SI{12.5}{\kelvin} and \SI{50}{\pico\second}, respectively.
We are able to perform these simulations by means of both granular model and atomistic calculations because the system is composed of an isolated single grain and consequently by only few thousands atoms.
Fig.~\ref{fig:phase_plot_FePt_atomistic_and_granular} shows the obtained phase plots for $\Hmax=\SI{0.5}{T}$ (a,c) and $\Hmax=\SI{1.0}{T}$ (b,d) comparing atomistic (top) and micromagnetic (bottom) simulations. 
The two different methods yield very similar results, as mentioned above, and hence we can use the LLB dynamics to perform more extensive calculations.
From these phase plots we can observe how shorter time pulses require higher peak temperatures to achieve a successful magnetisation reversal for a given external field. 
Similarly, stronger \Hmax needs to be applied for short \tpulse at a fixed temperature, which suggests the necessity for a trade-off between \Hmax, \tpulse and \Tpeak.  
A feature emerging from our results is that the magnetisation of a single grain of a HAMR granular medium can be reversed with probability larger than \SI{0.9}{}  only when the peak temperature is above \Tc and for strong applied fields. 

\section{Conclusions}
In summary, we have presented a multiscale approach that combines atomistic and micromagnetic simulations to model and describe HAMR media and their dynamics. 
The multiscale approach on one side exploits the high detail achievable throughout atomistic calculations and on the other uses this to provide the input parameters necessary for a micromagnetic model based on LLB dynamics. 
This method makes possible to overcome the computational limitations in dealing with large systems of atomistic simulations while retaining the high accuracy in the results. 
Our initial simulations prove the appropriateness and potential of the approach here proposed where the granular model is able to reproduce the atomistic simulations and main properties of a HAMR medium can be modelled. We show that careful atomistic parameterisation of the LLB equation is important in order to take into account the effects of finite grain size.
We have modelled the simple case of a magnetic layer composed of a granular FePt medium subjected to spatially uniform field and temperature pulses. The grain size of \SI{5}{nm} is smaller than current designs and represents an investigation of future HAMR media.  
Only switching probabilities obtained under the assumption of a \SI{1}{\tesla} field, the maximum likely for inductive technology, show good performances.
Therefore, alternatives such as magnetic layers made of exchange coupled composite (ECC) materials need to be pursued to make HAMR a viable technology. In addition, the magnetisation dynamics exhibits a mixture of precessional and linear character, differently from what is commonly assumed for HAMR processes. Our results suggest a strong dependence of the reversal mechanism on the properties of the temperature pulse. These aspects are crucial to improve HAMR technology and and will be the subject of future work.

\begin{acknowledgments}
JC would like to acknowledge the financial supported by Mahasarakham University. The authors gratefully acknowledge the funding from the Industry/ Academia Partnership Programme of the Royal Academy of Engineering and the Thailand Research Fund (grant IAPP1R2/100151) and Seagate Technology (Thailand). Financial support of the Advanced Storage Research consortium is gratefully acknowledged.
The authors would like to thank the York Advanced Computing Cluster (YARCC) for access to computational resources.
\end{acknowledgments}

%\bibliography{bibliography}
%merlin.mbs apsrev4-1.bst 2010-07-25 4.21a (PWD, AO, DPC) hacked
%Control: key (0)
%Control: author (8) initials jnrlst
%Control: editor formatted (1) identically to author
%Control: production of article title (-1) disabled
%Control: page (0) single
%Control: year (1) truncated
%Control: production of eprint (0) enabled
%

\end{document}